\documentclass[conference]{IEEEtran}
\IEEEoverridecommandlockouts
\usepackage{booktabs}

\usepackage{cite}
\usepackage{amsmath,amssymb,amsfonts}
\usepackage{algorithmic}
\usepackage{graphicx}
\usepackage{textcomp}
\usepackage{xcolor}
\usepackage{siunitx}
\usepackage{booktabs}
\usepackage{tabularx}
\usepackage{caption}
\usepackage{multirow}
\usepackage[numbers,sort&compress]{natbib}
\def\BibTeX{{\rm B\kern-.05em{\sc i\kern-.025em b}\kern-.08em
    T\kern-.1667em\lower.7ex\hbox{E}\kern-.125emX}}
\usepackage{cite}
\begin{document}

\title{Microelectrode Signal Dynamics as Biomarkers of Subthalamic Nucleus Entry on Deep Brain Stimulation: A Nonlinear Feature Approach}

\author{
\IEEEauthorblockN{1\textsuperscript{st} Ana Luiza Souza Tavares}
\IEEEauthorblockA{\textit{Institute of Biological Sciences} \\
\textit{Federal University of Pará} \\
Belém, PA, Brazil \\
ana.souza.tavares@icb.ufpa.br}
\and
\IEEEauthorblockN{2\textsuperscript{nd} Artur Pedro Martins Neto}
\IEEEauthorblockA{\textit{Institute of Biological Sciences} \\
\textit{Federal University of Pará} \\
Belém, PA, Brazil \\
artur.neto@icb.ufpa.br}
\and
\IEEEauthorblockN{3\textsuperscript{rd} Francinaldo Lobato Gomes}
\IEEEauthorblockA{\textit{Hospital Ophir Loyola} \\
Belém, PA, Brazil \\
francinaldogomes2012@hotmail.com}
\and
\IEEEauthorblockN{4\textsuperscript{th} Paul Rodrigo dos Reis}
\IEEEauthorblockA{\textit{Institute of Biological Sciences} \\
\textit{Federal University of Pará} \\
Belém, PA, Brazil \\
paulrodrigo@gmail.com}
\and
\IEEEauthorblockN{5\textsuperscript{th} Arthur Gonsales da Silva}
\IEEEauthorblockA{\textit{Faculty of Medical Sciences} \\
\textit{UNICAMP} \\
Campinas, SP, Brazil \\
aerthurg@posteo.com}
\and
\IEEEauthorblockN{6\textsuperscript{th} Antonio Pereira Junior}
\IEEEauthorblockA{\textit{Electrical Engineering Graduate Program} \\
\textit{Neuroscience and Cell Biology Graduate Program} \\
\textit{Federal University of Pará} \\
Belém, PA, Brazil \\
apereira@ufpa.br}
\and
\IEEEauthorblockN{7\textsuperscript{th} Bruno Duarte Gomes}
\IEEEauthorblockA{\textit{Electrical Engineering Graduate Program} \\
\textit{Neuroscience and Cell Biology Graduate Program} \\
\textit{Federal University of Pará} \\
Belém, PA, Brazil \\
brunodgomes@ufpa.br}
}

\maketitle

\begin{abstract}
Accurate intraoperative localization of the subthalamic nucleus (STN) is essential for the efficacy of Deep Brain Stimulation (DBS) in patients with Parkinson’s disease. While microelectrode recordings (MERs) provide rich electrophysiological information during DBS electrode implantation, current localization practices often rely on subjective interpretation of signal features. In this study, we propose a quantitative framework that leverages nonlinear dynamics and entropy-based metrics to classify neural activity recorded inside versus outside the STN. MER data from three patients were preprocessed using a robust artifact correction pipeline, segmented, and labeled based on surgical annotations. A comprehensive set of recurrence quantification analysis, nonlinear, and entropy features were extracted from each segment. Multiple supervised classifiers were trained on every combination of feature domains using stratified 10-fold cross-validation, followed by statistical comparison using paired Wilcoxon signed-rank tests with Holm–Bonferroni correction. The combination of entropy and nonlinear features yielded the highest discriminative power, and the Extra Trees classifier emerged as the best model with a cross-validated F1-score of 0.902$\pm$0.027 and ROC AUC of 0.887$\pm$0.055. Final evaluation on a 20\% hold-out test set confirmed robust generalization (F1 = 0.922, ROC AUC = 0.941). These results highlight the potential of nonlinear and entropy signal descriptors in supporting real-time, data-driven decision-making during DBS surgeries.
\end{abstract}


\section{Introduction}
Deep Brain Stimulation (DBS) of the subthalamic nucleus (STN) has become a cornerstone surgical treatment for advanced Parkinson’s disease (PD), effectively reducing motor symptoms and medication requirements by delivering high-frequency electrical pulses to modulate pathological basal ganglia circuits\citep{vinke2022}.

\begin{figure}[htbp]
    \centering
    \includegraphics[width=0.9\columnwidth]{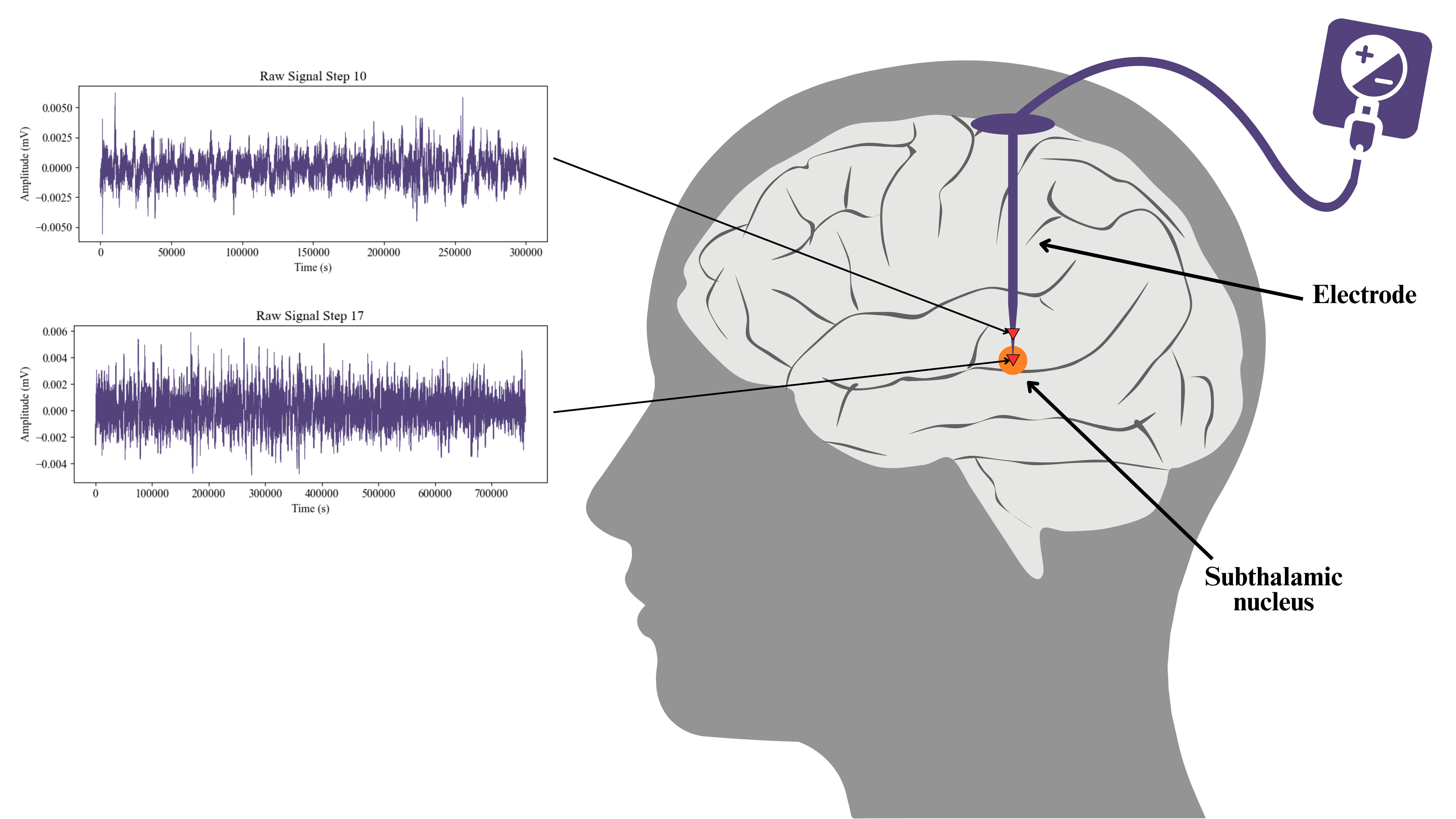}
    \caption{
\textbf{Schematic representation of DBS surgical procedure.}
The figure shows the DBS electrode targeting the STN, connected to the external stimulation and recording system. Red triangles mark the MER steps taken at 0.5 mm intervals during descent, enabling electrophysiological mapping to support accurate STN localization.
    }
    \label{fig:dbs_mer_schematic}
\end{figure}

Accurate STN electrode placement during DBS is critical for optimizing outcomes and minimizing side effects. Intraoperative microelectrode recordings (MERs) provide real-time electrophysiological guidance by detecting characteristic neuronal discharges as a high-impedance microelectrode advances along the surgical trajectory. This procedure is illustrated in Figure~\ref{fig:dbs_mer_schematic}, which shows the DBS system and representative MER recording steps along the descent path. Neurophysiologists monitor changes in spike frequency, amplitude, and background activity to delineate STN boundaries—particularly distinguishing transitions into the substantia nigra pars reticulata (SNr). This technique enables functional mapping of the subthalamic region and refines the permanent lead placement \citep{vinke2022}.
However, delineating the exact dorsal and ventral borders of the STN using MER can be challenging. The transition from preceding structures (e.g., thalamus or zona incerta) into the STN is often subtle, and distinguishing the STN from neighboring nuclei like the SNr requires expert interpretation of complex signal patterns \citep{karthick2020, hosny2021}. In current practice, identifying STN entry is largely performed manually by a neurophysiologist observing MER signals in real time \citep{vinke2022}. Furthermore, MER signals are both stochastic and non-stationary \citep{karthick2020}, meaning their statistical properties evolve with time and depth—further complicating consistent interpretation. These limitations motivate the need for objective, data-driven tools to assist in STN localization during DBS surgery.

Despite some progress in automating MER classification for DBS lead placement, significant challenges remain. Early approaches extracted hand-crafted spike-dependent and spike-independent features—such as firing rates, burst indices, and signal variance—to distinguish STN from non-STN regions \citep{karthick2020}. Wong et al. used a combination of these features within an unsupervised clustering framework to improve STN detection accuracy \citep{wong2009}. Similarly, Chaovalitwongse et al. and Ciecierski et al. applied classification algorithms using comparable feature sets, achieving moderate success \citep{chaovalitwongse2011, ciecierski2014}. However, most models relied on a limited subset of features and failed to comprehensively characterize the signal’s nonlinear dynamics, limiting robustness and interpretability.

Current MER analysis frameworks are inadequate in capturing the complex dynamics at STN boundaries, remaining heavily reliant on linear descriptors like spike counts or spectral energy, or simplified time-frequency analyses \citep{hosny2021}. While some studies explored entropy-based metrics, they often considered only one or two measures without integrating broader nonlinear descriptors. No existing models comprehensively combine entropy variants, fractal complexity, and chaos-theoretic measures to enhance localization accuracy.

To address this, nonlinear time-series techniques—particularly recurrence quantification analysis (RQA)—offer promising tools for modeling MER dynamics. RQA examines patterns of recurrence in phase-space trajectories, enabling detection of nonlinear state transitions without assuming signal stationarity \citep{zbilut1992, marwan2011}. Key RQA metrics include the recurrence rate (RR), average diagonal line length (L), and determinism (DET), which respectively reflect trajectory periodicity, temporal predictability, and structural stability during transitions through neural regions.

Beyond RQA, additional nonlinear metrics further quantify the complexity of neural signals. The largest Lyapunov exponent measures exponential divergence of trajectories and reveals chaotic dynamics \citep{rosenstein1993}. The Hurst exponent estimates long-term memory and autocorrelation \citep{hurst1951}. Fractal dimensions such as Higuchi’s and Katz’s capture signal roughness and geometric complexity \citep{higuchi1988, katz1988}. Lempel–Ziv complexity (LZC) quantifies sequence randomness in binarized signals and is widely used as a model-free complexity measure \citep{lempel1976}.

Entropy-based metrics also provide critical insights into signal irregularity and information content. Shannon entropy captures the overall uncertainty of amplitude distributions \citep{shannon1948}. Permutation entropy evaluates local ordinal pattern diversity \citep{bandt2002}, while sample entropy (SampEn) and approximate entropy (ApEn) assess the probability and regularity of repeated temporal patterns, respectively \citep{richman2000, pincus1991}. Tsallis entropy generalizes Shannon’s framework to better handle non-extensive statistical systems and long-range dependencies \citep{tsallis1988}. Together, these nonlinear and entropic measures offer a powerful toolkit for characterizing electrophysiological complexity within and beyond the STN.

This study aims to develop a supervised, data-driven MER classification framework that leverages a comprehensive suite of nonlinear and entropy-based features to distinguish STN from adjacent regions. By quantifying irregularity, complexity, and chaos in the MER signal—markers that may reflect functional transitions during electrode descent—we hypothesize that recordings from within the STN exhibit distinct nonlinear profiles. Our goal is to demonstrate that integrating these features can serve as a robust biomarker system for intraoperative STN localization, enhancing objectivity and accuracy in DBS lead placement.

\section{METHODOLOGY}
\subsection{DBS Dataset}\label{AA}
Intraoperative MERs were analyzed from three idiopathic Parkinson's disease patients undergoing STN DBS implantation under local anesthesia. Institutional review board approval and written informed consent were obtained. Two independent electrode descents per hemisphere followed stereotactically defined trajectories, beginning 10 mm above the radiologically determined STN dorsal border and proceeding in 0.5 mm increments to 2 mm below the ventral border. Signals were digitized at 20 kHz with 16-bit resolution and stored in EDF format. Two patients underwent dual-channel MER (parallel trajectories separated by 2 mm). Real-time annotation of each step as "outside STN" or "inside STN" was performed by an experienced neurophysiologist based on established electrophysiological hallmarks. All recordings and metadata were anonymized and organized in a secure database.

\begin{figure*}[htbp]
    \centering
    \includegraphics[width=0.8\textwidth]{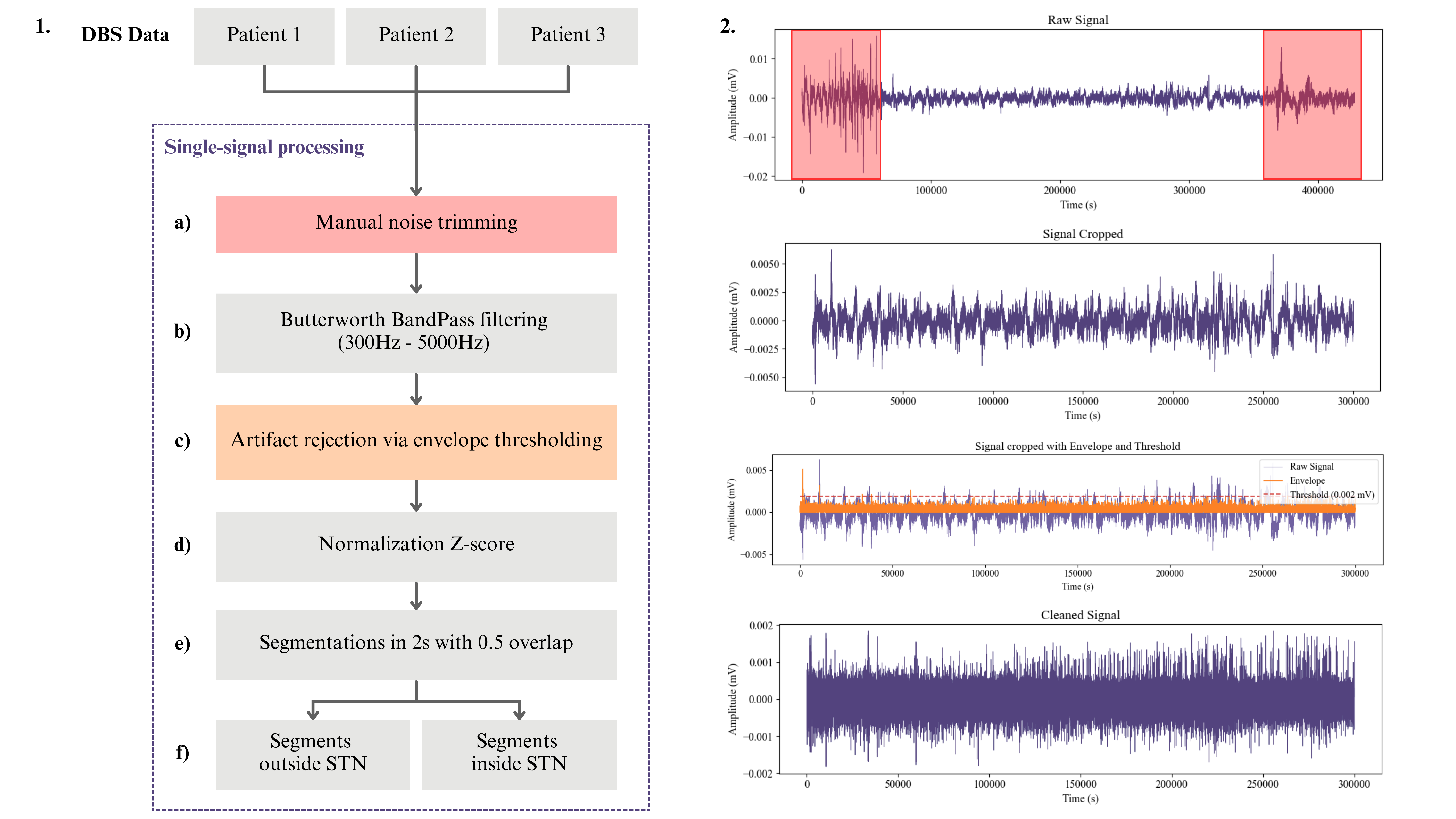}
    \caption{
\textbf{Overview of the preprocessing pipeline for MER signals.} 
\textbf{(Item 1)} Flowchart of the complete preprocessing procedure, illustrating each processing step applied to the raw microelectrode recordings of each patient: 
(a) manual noise trimming; 
(b) Butterworth bandpass filtering (300–5000 Hz); 
(c) artifact rejection via envelope thresholding; 
(d) Z-score normalization; 
(e) segmentation into 2-second overlapping windows; 
(f) anatomical labeling as inside or outside the STN. 
\textbf{(Item 2)} Representative signal snapshots corresponding to the artifact rejection stage: 
(i) raw signal with manual crop markers; 
(ii) cropped signal after noise removal; 
(iii) cropped signal with overlaid Hilbert envelope and adaptive threshold; 
(iv) cleaned signal following envelope-based artifact rejection and interpolation PCHIP.
}
    \label{processing}
\end{figure*}

\subsection{Signal Preprocessing Pipeline}
To ensure the integrity and analytical reliability of raw MER, all data was subjected to a rigorous preprocessing pipeline. This multistage procedure was implemented in Python and was designed to systematically reduce mechanical artifacts, remove high-amplitude transient disturbances, standardize signal properties, and prepare the data for robust feature extraction. Figure~\ref{processing} illustrates this preprocessing pipeline, presenting both a flowchart of the full sequence and representative signal snapshots from the artifact rejection process. Each step was informed by established practices in neural signal analysis and adapted to the characteristics of intraoperative MER acquired during the implantation of the DBS electrode.

Raw MER segments underwent preprocessing starting with manual artifact trimming via a Python interface, where transient mechanical noise at recording onset/offset was excised through visual time-domain inspection and cropping of artifact-free intervals. Subsequently, signals underwent zero-phase bandpass filtering (300-5000 Hz) using a 4th-order Butterworth configuration—a standard MER processing range that preserves spike morphology by eliminating phase distortion while attenuating low-frequency movement artifacts and high-frequency electronic noise.

Non-physiological transients were mitigated via adaptive envelope thresholding adapted from intraoperative denoising frameworks. Pre-filtered signals (300-5000 Hz) underwent Hilbert-derived envelope extraction, with baseline noise estimated through modal histogram analysis refined via local polynomial fitting. An adaptive 8$\sigma$ threshold flagged artifact intervals, which were reconstructed via Piecewise Cubic Hermite Interpolation (PCHIP) to preserve intrinsic spike dynamics while eliminating transient distortions \citep{schiff2009}.

Following artifact correction, signal segments underwent z-score normalization to standardize amplitude scales across diverse recording sources (patients, trajectories, channels), ensuring cross-comparability. Preprocessed data was then partitioned into 2-second windows with 0.5-second overlaps, balancing temporal resolution with neural dynamic continuity while augmenting dataset granularity. Each window inherited binary anatomical labels ("inside STN" or "outside STN") from intraoperative neurophysiological annotations, forming the ground truth for subsequent supervised classification.

\begin{table}[htbp]
\centering
\footnotesize
\caption{Summary of feature groups and their computed metrics.}
\setlength{\tabcolsep}{20pt}
\label{tab:metrics}
\begin{tabular*}{\columnwidth}{@{\extracolsep{\fill}} p{0.30\columnwidth} p{0.70\columnwidth} @{}}
\toprule
\textbf{Feature Group} & \textbf{Metrics} \\
\midrule
RQA & Recurrence Rate (RR)\\
 &  Determinism (DET)\\
 &  Average Diagonal Line Length (L) \\
\midrule
Nonlinear Dynamics & Largest Lyapunov Exponent ($\lambda_1$)\\
 & Hurst Exponent (H)\\
 &  Higuchi Fractal Dimension (HFD)\\
 &  Katz Fractal Dimension\\
 &  Lempel–Ziv Complexity (LZC) \\
\midrule
Entropic Measures & Shannon Entropy (ENT)\\
 &  Permutation Entropy\\
 &  Sample Entropy (SampEn)\\
 &  Approximate Entropy (ApEn)\\
 & Tsallis Entropy \\
\bottomrule
\end{tabular*}
\end{table}

\subsection{Feature Extraction}

To comprehensively characterize the dynamical properties of MERs, a broad suite of nonlinear and entropy-based features was extracted from each 2-second segment. These features were organized into three main domains: Recurrence Quantification Analysis (RQA), nonlinear dynamical metrics, and entropy-based complexity measures. All metrics were computed using established Python packages and extracted independently for each segment.

\subsubsection{Recurrence Quantification Analysis (RQA)}

RQA is a nonlinear time-series analysis technique that assesses the recurrence of states in a reconstructed phase space. The phase space is reconstructed from the original time series \( x(t) \) using time-delay embedding with embedding dimension \( m \) and delay \( \tau \). The recurrence plot is defined as:

\begin{equation}
R_{i,j} = \Theta(\epsilon - \| \mathbf{x}_i - \mathbf{x}_j \|),
\end{equation}

where \( \Theta(\cdot) \) is the Heaviside step function, \( \epsilon \) is a recurrence threshold, and \( \| \cdot \| \) denotes the Euclidean norm.

From the recurrence plot, the following RQA metrics were computed:

\begin{itemize}
    \item {Recurrence Rate (RR):} the proportion of recurrence points in the matrix:
    \begin{equation}
    RR = \frac{1}{N^2} \sum_{i,j=1}^{N} R_{i,j}
    \end{equation}

    \item {Average Diagonal Line Length (L):} the mean length of diagonal lines (i.e., periods of similar system evolution):
    \begin{equation}
    L = \frac{ \sum_{l = l_{\min}}^{N} l \cdot P(l) }{ \sum_{l = l_{\min}}^{N} P(l) }
    \end{equation}

    \item {Determinism (DET):} the fraction of recurrence points that form diagonal lines of minimum length:
    \begin{equation}
    DET = \frac{ \sum_{l = l_{\min}}^{N} l \cdot P(l) }{ \sum_{i,j=1}^{N} R_{i,j} }
    \end{equation}
\end{itemize}

\subsubsection{Nonlinear Dynamical Metrics}

These metrics provide insight into the chaotic, fractal, and memory-related features of the signals.

\begin{itemize}
    \item {Largest Lyapunov Exponent (LLE):} computed using Rosenstein’s algorithm, this exponent quantifies sensitivity to initial conditions:
    \begin{equation}
    \lambda = \lim_{t \to \infty} \frac{1}{t} \ln \left( \frac{d(t)}{d(0)} \right)
    \end{equation}

    \item {Hurst Exponent (H):} indicates long-range memory:
    \begin{equation}
    E[R(n)/S(n)] \propto n^H
    \end{equation}

    \item {Higuchi Fractal Dimension (HFD):} estimates signal complexity based on how curve length varies with scale \( k \):
    \begin{equation}
    D = \lim_{k \to 0} \frac{\log L(k)}{\log(1/k)}
    \end{equation}

    \item {Katz Fractal Dimension (KFD):} another fractal measure based on the total signal length \( L \) and diameter \( d \):
    \begin{equation}
    D = \frac{\log(n)}{\log(n) + \log(d/L)}
    \end{equation}

    \item {Lempel–Ziv Complexity (LZC):} counts unique patterns in a binarized sequence (median-thresholded):
    \begin{equation}
    LZC = \frac{c(n)}{n / \log_2 n}
    \end{equation}
\end{itemize}

\subsubsection{Entropy-Based Complexity Measures}

Entropy metrics quantify unpredictability and structural disorder in the time series.
\begin{figure*}[htbp]
    \centering
    \includegraphics[width=\textwidth]{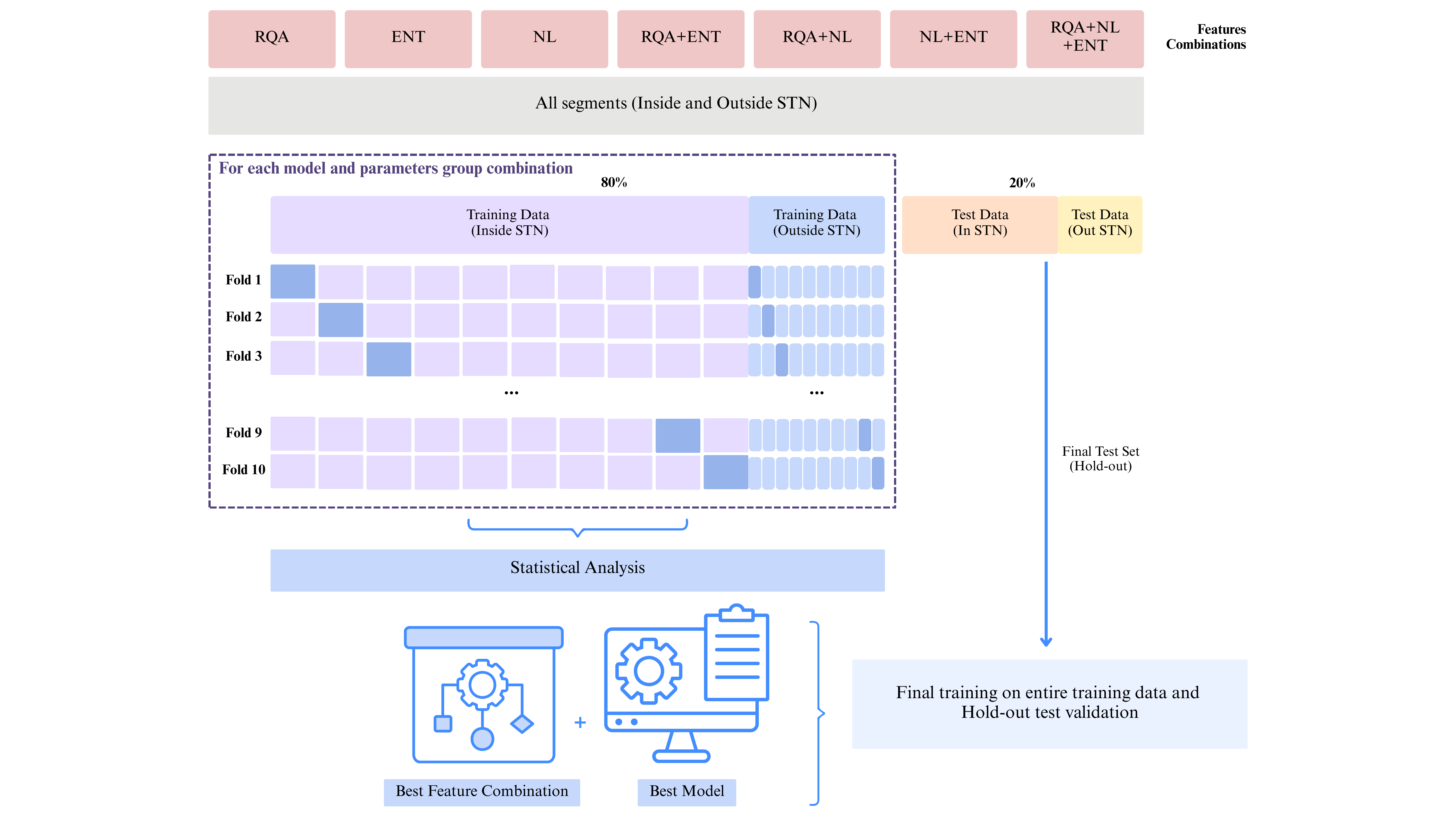}
    \caption{
\textbf{Overview of data partitioning, feature-group evaluation, and final model selection.} 
All labeled MER signal segments were first aggregated and used to construct a feature matrix spanning three domains: RQA, Entropy-based metrics (ENT), and Nonlinear dynamics (NL). All non-empty combinations of these three domains were enumerated, resulting in seven distinct feature-group configurations for performance assessment (top row). The full dataset was split using stratified random sampling into a training set (80\%) and a final hold-out test set (20\%).
Within the training set, 10-fold stratified cross-validation was performed independently for each combination of feature group and classification model. For each fold, classifier performance was evaluated using F1-score, Accuracy, Recall, Precision and ROC-AUC. Statistical comparison across folds was used to identify the most informative feature group and the best-performing classifier using Wilcoxon signed-rank tests. Finally, the model with the highest statistically supported performance was retrained on the full training data and evaluated on the untouched hold-out test set intervals to assess generalization.
}
    \label{fig:yourlabel}
\end{figure*}
\begin{itemize}
    \item {Shannon Entropy:} quantifies the uncertainty in the signal amplitude distribution:
    \begin{equation}
    H = - \sum_{i=1}^{N} p_i \log_2 p_i
    \end{equation}

    \item {Permutation Entropy (PE):} measures the diversity of ordinal patterns:
    \begin{equation}
    H_{PE} = - \sum_{j=1}^{m!} p_j \log_2 p_j
    \end{equation}

    \item {Sample Entropy (SampEn):} evaluates the regularity of repeated patterns of length \( m \):
    \begin{equation}
    \text{SampEn}(m,r) = -\ln \left( \frac{A}{B} \right)
    \end{equation}

    \item {Approximate Entropy (ApEn):} assesses predictability by comparing patterns of lengths \( m \) and \( m+1 \):
    \begin{equation}
    \text{ApEn}(m,r) = \Phi^m(r) - \Phi^{m+1}(r)
    \end{equation}
    \begin{equation}
    \Phi^m(r) = \frac{1}{N - m + 1} \sum_{i=1}^{N-m+1} \log C_i^m(r)
    \end{equation}

    \item {Tsallis Entropy:} a generalized entropy measure for non-extensive systems:
    \begin{equation}
    S_q = \frac{1}{q-1} \left( 1 - \sum_{i=1}^{N} p_i^q \right)
    \end{equation}
\end{itemize}

This diverse set of nonlinear and entropy-based features was extracted from each signal segment to support fine-grained classification of MERs based on subcortical localization. Together, these metrics provide a multidimensional representation of signal dynamics, enabling the identification of subtle but informative differences in neural activity patterns across anatomical boundaries.

\subsection{Classification Framework}
To systematically evaluate the discriminative power of the RQA, nonlinear and entropy‐based features for STN localization, we implemented a multi‐stage classification protocol comprising stratified cross‐validation, feature‐group evaluation and statistical selection, and final hold‐out testing. An overview of the classification architecture—including data partitioning, feature-group enumeration, model evaluation, and statistical selection—is summarized in Figure~\ref{fig:yourlabel}.

\subsubsection{Data Splitting and Feature‐Group Evaluation}
All preprocessed 2-second MER segments—each labeled “inside STN” or “outside STN”—were aggregated into a feature matrix comprising three domains: RQA, Entropy and Nonlinear features, as described in the previous section. The full dataset was stratified by label and split into a training set (80\%) and an independent hold‐out test set (20\%) using stratified random sampling.

To assess the classification power of each domain, we enumerated every non-empty combination of the three feature groups, yielding seven distinct representations: RQA alone, Entropy alone, Nonlinear alone, RQA+Entropy (R+E), RQA+Nonlinear (R+N), Entropy+Nonlinear (E+N) and RQA+Entropy+Nonlinear (R+E+N).

\subsubsection{Candidate Classifiers}

Seven classifiers were chosen to span complementary algorithmic families under comparison. In the tree‐based ensemble category, we evaluated Random Forest and Extra Trees, alongside a single Decision Tree. A gradient‐boosted alternative was represented by XGBoost. To capture nonlinear boundaries, we included a kernel method, namely an RBF‐kernel SVM with probability calibration and balanced weighting. An instance‐based approach was provided by k‐Nearest Neighbors, and finally, a probabilistic model was embodied by Gaussian Naïve Bayes (assuming feature‐wise Gaussian likelihoods). All scale‐sensitive algorithms were wrapped in a pipeline that applied Z‐score normalization to the selected features prior to training.

\subsubsection{Cross‐Validation and Performance Metrics}

Within the training set, we conducted stratified 10-fold cross‐validation. For each fold, classifier, and feature‐group configuration, the following metrics were computed on held‐out folds: Accuracy, Recall, F1-Score, Area under the ROC Curve.
\subsubsection{Statistical Model Selection}
To rigorously identify the optimal feature‐group permutation and classifier, we applied a two-stage paired Wilcoxon signed‐rank testing procedure (two-sided, $\alpha=0.05$), with Holm-Bonferroni correction for multiple comparisons to control the family-wise error rate.

\paragraph{Feature‐Group Comparison Across Classifiers}  
We first aggregated all models’ per-fold F1 and ROC-AUC values by feature‐group combination. The combination exhibiting the highest mean across these metrics was designated as the provisional best representation. To verify that its superiority was not due to chance, we performed paired Wilcoxon signed‐rank tests comparing the fold-wise performance of this reference group against each alternative combination. For each test we report the signed‐rank statistic $W$, the corrected two-sided $p$-value, and denote statistical significance ($p_{\text{corr}} < 0.05$).

\paragraph{Classifier Comparison Within Best Feature Group}  
Fixing the selected feature‐group representation, we extracted per-fold F1 and ROC-AUC scores for each classifier. The classifier with the highest average performance served as the reference model. We then conducted paired Wilcoxon signed-rank tests comparing its fold-wise scores against those of every other candidate, again applying Holm-Bonferroni correction to the $p$-values. We report $W$, corrected $p$-values, and significance indicators to confirm the statistical robustness of the selected classifier’s advantage.

\subsubsection{Final Model Training and Hold-Out Evaluation}

The classifier–feature‐group pairing that passed both statistical selection stages was retrained on the entire 80\% training set. Its generalization performance was then evaluated on the untouched 20\% hold-out test set, using Accuracy, Precision, Recall, F1-Score and ROC-AUC as primary evaluation metrics. To quantify the uncertainty associated with these metrics, we employed non-parametric bootstrapping with 1,000 resamples of the test set, computing 95\% confidence intervals for each metric.

\section{RESULTS}

\subsection{Data \& Preprocessing Overview}

Across all three patients and both cirurgy, the segmentation process produced 3463 analysis windows in total. Of these, 2526 windows (72.94\%) were labeled “outside STN” and 937 windows (27.06\%) “outside STN”.

Our preprocessing pipeline did not remove any windows outright. Instead, only 0.09\% of all data points were treated as artifacts and interpolated, preserving the full dataset size while correcting transient disturbances.

\subsection{Feature Summaries}

Table~\ref{tab:feature_summaries} reports the mean $\pm$ standard deviation (SD) values of the four features exhibiting the greatest separation between segments recorded inside versus outside the STN. The average diagonal line length (\(L\)) was substantially higher within the STN (4.089 $\pm$ 0.550) than outside (3.600 $\pm$ 0.451), reflecting more prolonged recurrences of similar dynamic states in the target region. Shannon entropy likewise increased from 2.141 $\pm$ 0.258 outside the STN to 2.339 $\pm$ 0.257 inside, indicating a broader amplitude distribution in intra‐nuclear recordings. In contrast, sample entropy decreased from 1.065 $\pm$ 0.122 outside to 0.917 $\pm$ 0.121 inside, suggesting that, despite greater overall amplitude variability, successive temporal patterns were more predictable within the STN. Finally, the Katz fractal dimension showed a modest reduction inside the nucleus (3.709 $\pm$ 0.220) compared to outside (3.843 $\pm$ 0.292), consistent with a slight decrease in signal geometric complexity at the target site.

\begin{table}[!htbp]
  \centering
  \caption{Mean $\pm$ SD of selected features showing the largest class separation between “inside STN” and “outside STN” segments.}
  \label{tab:feature_summaries}
  \begin{tabular}{lcc}
    \toprule
    \textbf{Metric}                               & \textbf{Inside STN} & \textbf{Outside STN} \\
                                                  & (mean $\pm$ SD)    & (mean $\pm$ SD)      \\
    \midrule
    Average Diagonal Line Length           & 4.089 $\pm$ 0.550  & 3.600 $\pm$ 0.451    \\
    Shannon Entropy                      & 2.339 $\pm$ 0.257  & 2.141 $\pm$ 0.258    \\
    Sample Entropy                    & 0.917 $\pm$ 0.121  & 1.065 $\pm$ 0.122    \\
    Katz Fractal Dimension          & 3.709 $\pm$ 0.220  & 3.843 $\pm$ 0.292    \\
    \bottomrule
  \end{tabular}
\end{table}

\subsection{Classification Performance}

\subsubsection{Cross-Validation Feature Groups Results}  
\begin{table}[!htbp]
\centering
\small
\caption{Cross‐validation performance (mean\,$\pm$\,SD) for each feature‐group combination.}
\label{tab:cv_results}
\resizebox{\linewidth}{!}{%
\begin{tabular}{lccc}
\hline
\textbf{Feature Group}               & \textbf{Recall}       & \textbf{F1‐Score}     & \textbf{AUC}          \\
\hline
Entropy + Nonlinear                  & $0.915\pm0.050$       & $0.886\pm0.034$       & $0.849\pm0.075$       \\
RQA + Entropy + Nonlinear            & $0.909\pm0.056$       & $0.882\pm0.036$       & $0.844\pm0.078$       \\
Entropy                              & $0.904\pm0.040$       & $0.880\pm0.028$       & $0.833\pm0.069$       \\
RQA + Entropy                        & $0.901\pm0.046$       & $0.877\pm0.029$       & $0.836\pm0.070$       \\
RQA + Nonlinear                      & $0.901\pm0.051$       & $0.869\pm0.033$       & $0.811\pm0.078$       \\
Nonlinear                            & $0.913\pm0.053$       & $0.863\pm0.029$       & $0.788\pm0.075$       \\
RQA                                  & $0.882\pm0.049$       & $0.843\pm0.023$       & $0.747\pm0.056$       \\
\hline
\end{tabular}
}
\end{table}

Table~\ref{tab:cv_results} reports mean $\pm$ SD for Recall, F1-Score and AUC across stratified 10-fold CV of seven classifiers. The Entropy+Nonlinear (E+N) set achieved the top performance (F1 = $0.886\pm0.034$; AUC = $0.849\pm0.075$), followed by RQA+Entropy+Nonlinear (R+E+N; F1 = $0.882\pm0.036$; AUC = $0.844\pm0.078$).

\begin{table*}[!htbp]
  \centering
  \scriptsize
  \setlength{\tabcolsep}{2pt}
  \caption{Paired Wilcoxon signed‐rank tests comparing Entropy+Nonlinear (E+N) against alternate feature groups.}
  \label{tab:stat_model_select}
  \begin{tabular*}{\textwidth}{@{\extracolsep{\fill}} l c c c c c @{}}
    \toprule
    \textbf{Feature Group} 
      & \textbf{ROC\_AUC} $W (p)$ 
      & \textbf{Accuracy} $W (p)$ 
      & \textbf{Precision} $W (p)$ 
      & \textbf{Recall} $W (p)$ 
      & \textbf{F1} $W (p)$ \\
    \midrule
    Entropy         & 399 (0.0000*)   & 775.5 (0.0151*) & 1127 (0.4991)   & 670.5 (0.0021*) & 797 (0.0091*)  \\
    Nonlinear       &   0 (0.0000*)   &  20 (0.0000*)   &    0 (0.0000*)  & 1001.5 (0.6426) & 119 (0.0000*)  \\
    RQA             &   1 (0.0000*)   &  76 (0.0000*)   &    2 (0.0000*)  & 461 (0.0000*)   & 132 (0.0000*)  \\
    RQA+Entropy     & 462 (0.0000*)   & 398 (0.0000*)   &  902 (0.0463*)  & 290.5 (0.0000*) & 445 (0.0000*)  \\
    RQA+Nonlinear   &   3 (0.0000*)   &  84 (0.0000*)   &   51.5 (0.0000*) & 367 (0.0000*)   & 123 (0.0000*)  \\
    RQA+E+N         & 825.5 (0.0147*) & 733.5 (0.0264*) & 1067 (0.4009)   & 510.5 (0.0018*) & 776 (0.0099*)  \\
    \bottomrule
    \multicolumn{6}{@{}l}{\scriptsize *$p<0.05$ }  
  \end{tabular*}
\end{table*}

Notably, the three feature domains exhibit markedly different standalone strengths: Entropy alone attains strong discrimination (F1 = $0.880\pm0.028$; AUC = $0.833\pm0.069$), outperforming both Nonlinear (F1 = $0.863\pm0.029$; AUC = $0.788\pm0.075$) and RQA (F1 = $0.843\pm0.023$; AUC = $0.747\pm0.056$).

To verify that these observed differences were not due to chance, we conducted paired Wilcoxon signed‐rank tests comparing E+N against each alternative. The statistical results are shown in Table~\ref{tab:stat_model_select}. These tests confirm that E+N significantly outperforms all other combinations on AUC and F1 ($p_{\mathrm{corr}}<0.05$), and on Accuracy except versus Entropy alone ($p_{\mathrm{corr}}=0.0151$). By contrast, E+N and Entropy do not differ significantly in Precision ($p_{\mathrm{corr}}=0.4991$), nor do E+N and Nonlinear in Recall ($p_{\mathrm{corr}}=0.6426$).

These results reinforce that entropy features capture the bulk of class‐separable information, and that selective augmentation with nonlinear descriptors yields a modest but significant gain (E+N vs. Entropy: $\Delta$F1 = 0.006; $\Delta$AUC = 0.016). Conversely, inclusion of RQA metrics offers no additional benefit—and indeed increases variability (R+E+N vs. E+N: $\Delta$F1 = –0.004; $\Delta$AUC = –0.005; higher AUC SD)—indicating redundancy with fractal measures and sensitivity to threshold selection. In sum, the statistically validated optimum is the Entropy+Nonlinear combination, which maximizes performance while avoiding superfluous dimensions.

\subsubsection{Statistical Model Selection}  
Within the E+N feature space, seven classifiers were evaluated via the stratified 10-fold CV. Table~\ref{tab:model_metrics} reports their mean ± SD F1-Score and ROC AUC. Extra Trees achieved the highest performance (F1 = $0.902\pm0.027$, AUC = $0.887\pm0.055$).  

\begin{figure}[!htbp]
  \centering
  \includegraphics[width=0.75\linewidth]{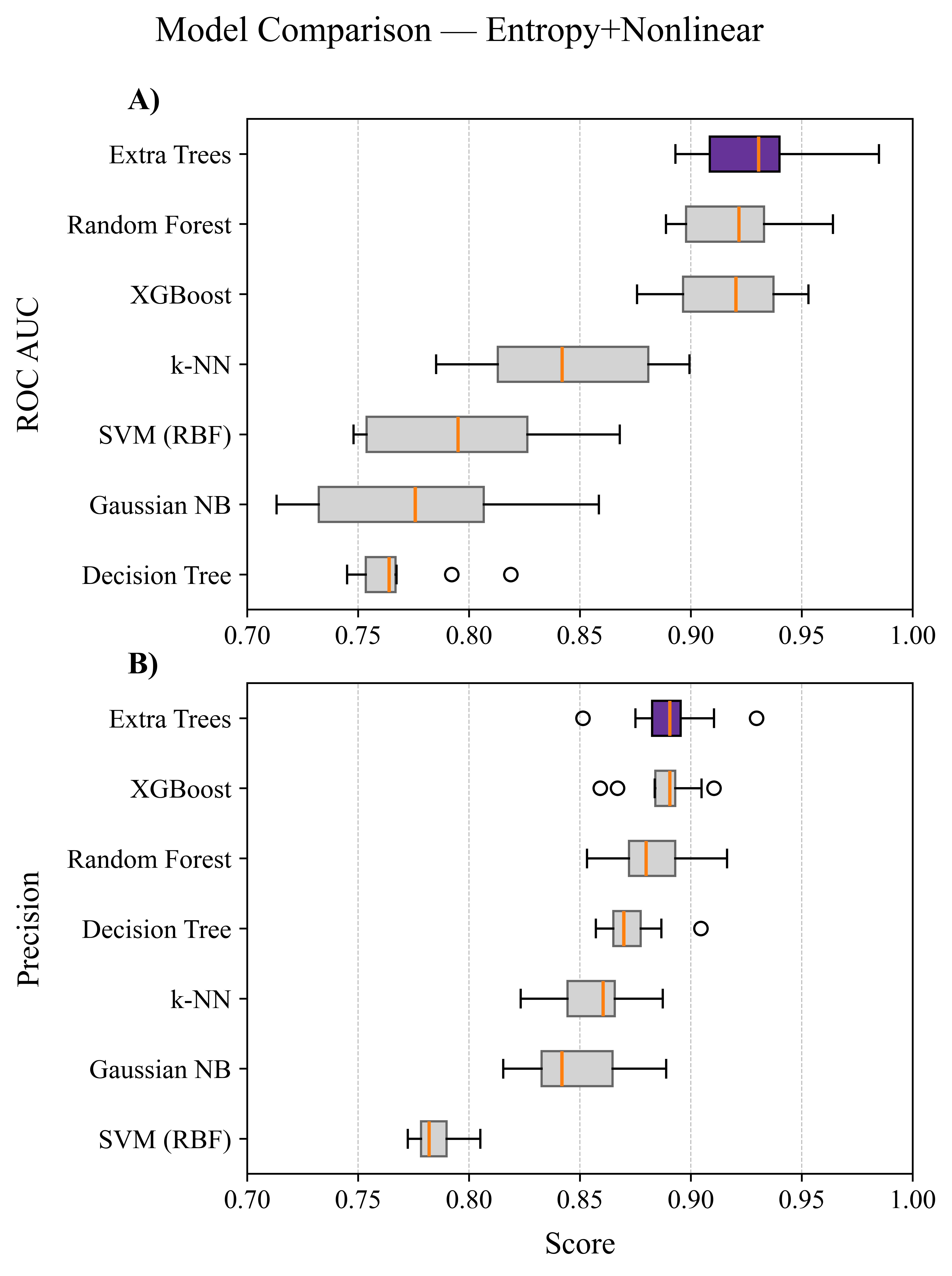}
  \caption{Boxplot of per-fold ROC AUC (A) and Precision (B) for each classifier within the E+N feature set. Extra Trees exhibits superior performance.}
  \label{fig:model_auc_boxplot}
\end{figure}

Fig.~\ref{fig:model_auc_boxplot} illustrates the per-fold ROC AUC distributions for each classifier. Also, Paired Wilcoxon signed-rank tests (Holm–Bonferroni-corrected, $\alpha=0.05$) confirm that Extra Trees significantly outperforms all alternatives on ROC AUC and F1 (all $p_{\mathrm{corr}}<0.05$). Examination of the remaining metrics revealed that Extra Trees’ advantage in Precision was not significant only versus XGBoost ($p_{\mathrm{corr}}=0.4922$), and in Recall was not significant only versus SVM (RBF) ($p_{\mathrm{corr}}=0.1953$).

These results establish Extra Trees as the statistically superior model in the Entropy+Nonlinear feature space.

\begin{table}[!htbp]
\centering
\caption{Cross-validated F1-Score and ROC AUC of classifiers within the Entropy+Nonlinear feature set (mean ± SD).}
\label{tab:model_metrics}
\begin{tabular}{lcc}
\toprule
\textbf{Classifier}        & \textbf{F1-Score}      & \textbf{ROC\_AUC}      \\
\midrule
Extra Trees                & $0.902\pm0.027$        & $0.887\pm0.055$        \\
Random Forest              & $0.898\pm0.025$        & $0.881\pm0.050$        \\
XGBoost                    & $0.890\pm0.025$        & $0.873\pm0.051$        \\
k-Nearest Neighbors (k-NN) & $0.860\pm0.025$        & $0.807\pm0.055$        \\
SVM (RBF kernel)           & $0.857\pm0.013$        & $0.768\pm0.053$        \\
Gaussian Naïve Bayes       & $0.842\pm0.026$        & $0.765\pm0.049$        \\
Decision Tree              & $0.852\pm0.028$        & $0.726\pm0.045$        \\
\bottomrule
\end{tabular}
\end{table}
\begin{table}[!htbp]
\centering
\caption{Hold‐out test performance of the Extra Trees model (mean and 95 \% bootstrap CIs).}
\label{tab:holdout_results}
\begin{tabular}{lccc}
\toprule
\textbf{Metric}    & \textbf{Mean} & \textbf{2.5 \% CI} & \textbf{97.5 \% CI} \\
\midrule
Accuracy           & 0.882         & 0.859             & 0.905              \\
Precision          & 0.884         & 0.858             & 0.910              \\
Recall             & 0.964         & 0.947             & 0.980              \\
F1‐score           & 0.922         & 0.906             & 0.938              \\
ROC AUC            & 0.941         & 0.922             & 0.957              \\
\bottomrule
\end{tabular}
\end{table}
\subsubsection{Final Hold‐Out Evaluation}  
\begin{figure}[!htbp]
  \centering
  \includegraphics[width=0.8\linewidth]{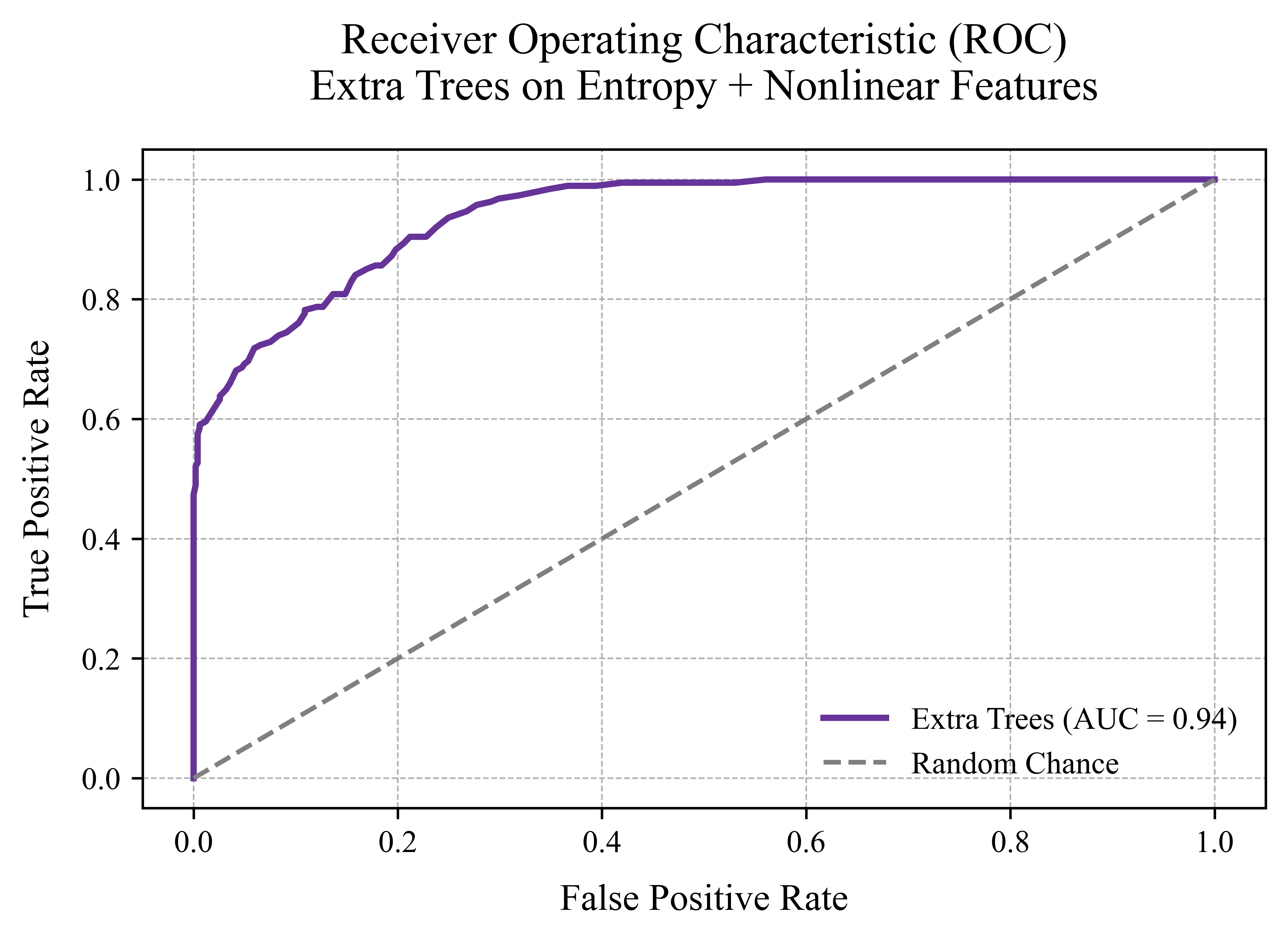}
  \caption{ROC curve for the Extra Trees classifier trained on the Entropy+Nonlinear feature set. The solid purple line plots true positive rate versus false positive rate on the hold-out test set, yielding an AUC of 0.92. The dashed gray diagonal denotes random-chance performance (AUC = 0.50).}
  \label{fig:roc_curve}
\end{figure}

Finally, we trained the best performing classifier, Extra Trees on the full 80\% training set with the best performing feature group, Entropy+Nonlinear, and evaluated on the independent 20\% hold‐out test set. Table~\ref{tab:holdout_results} presents the mean performance metrics along with their 95 \% bootstrap confidence intervals (2.5 \% and 97.5 \% percentiles). Fig.~\ref{fig:roc_curve} display the ROC curve (with AUC) for further presentation of model behavior on unseen data.

\section{CONCLUSION}
In this study, we have demonstrated that a careful combination of entropy‐based and nonlinear dynamical features provides a robust signature for distinguishing “inside” from “outside” STN microelectrode recordings. Our feature‐level analysis revealed that recurrence patterns (average diagonal line length), amplitude complexity (Shannon entropy), temporal predictability (sample entropy) and geometric structure (Katz fractal dimension) each contribute uniquely to class separation, underscoring their biophysical relevance to STN boundaries. When pooled, these metrics in the Entropy+Nonlinear feature set consistently outperformed alternate domain combinations, and a tree‐ensemble classifier delivered state‐of‐the‐art accuracy.

The Extra Trees model offers transparent insight into which signal properties drive localization decisions, an important advantage for intraoperative guidance. Moreover, rigorous statistical validation (paired Wilcoxon tests) and reliable generalization on an independent hold‐out test set confirm that our framework is not only accurate but also stable across folds and patients. By avoiding redundant or variance‐inflating feature groups (such as RQA), the pipeline remains computationally efficient, paving the way for real‐time deployment.

Future efforts will prioritize integrating this algorithmic workflow into the surgical environment and validating it on larger, multicenter datasets will be critical next steps. The modular nature of our preprocessing, feature‐extraction and classification components facilitates extension to additional signal modalities (e.g., local field potentials or multi‐site recordings) and potential adaptation to other DBS targets. Ultimately, this data‐driven approach holds promise for enhancing the precision and safety of DBS electrode placement through real‐time, interpretable neural‐signal analysis.


\begin{thebibliography}{99}

\bibitem{vinke2022} R.~S. Vinke \emph{et al.}, ``The role of microelectrode recording in deep brain stimulation surgery for Parkinson’s disease: a systematic review and meta-analysis,'' \emph{J. Parkinson’s Dis.}, vol.~12, no.~7, pp. 2059--2069, 2022.

\bibitem{karthick2020} P.~A. Karthick \emph{et al.}, ``Automated detection of subthalamic nucleus in deep brain stimulation surgery for Parkinson’s disease using microelectrode recordings and wavelet packet features,'' \emph{J. Neurosci. Methods}, vol.~343, p. 108826, 2020.

\bibitem{hosny2021} M. Hosny \emph{et al.}, ``Deep convolutional neural network for the automated detection of subthalamic nucleus using MER signals,'' \emph{J. Neurosci. Methods}, vol.~356, p. 109145, 2021.

\bibitem{wong2009} K. Wong \emph{et al.}, ``Unsupervised clustering of subthalamic nucleus MER features for DBS targeting,'' \emph{J. Neural Eng.}, vol.~6, no.~4, p. 046003, 2009.

\bibitem{chaovalitwongse2011} M. Chaovalitwongse \emph{et al.}, ``Automated classification of microelectrode recordings in deep brain stimulation surgery,'' \emph{IEEE Trans. Biomed. Eng.}, vol.~58, no.~6, pp. 1786--1794, 2011.

\bibitem{ciecierski2014} E. Ciecierski \emph{et al.}, ``Machine learning for intraoperative mapping of the subthalamic nucleus,'' \emph{Front. Neurol.}, vol.~5, Art. 183, 2014.

\bibitem{zbilut1992} J.~P. Zbilut and C.~L. Webber, ``Recurrence quantification analysis,'' \emph{Phys. Lett. A}, vol.~171, no.~3--4, pp. 199--203, 1992.

\bibitem{marwan2011} N. Marwan \emph{et al.}, ``Complex network analysis of recurrence networks,'' \emph{Eur. Phys. J. B}, vol.~84, no.~4, pp. 635--653, 2011.

\bibitem{cofre2025} R. Cofré and A. Destexhe, ``Entropy and complexity tools across scales in neuroscience: a review,'' \emph{Entropy}, vol.~27, no.~2, Art. 115, 2025.

\bibitem{rosenstein1993} M.~T. Rosenstein, J.~J. Collins, and C.~J. De Luca, ``A practical method for calculating largest Lyapunov exponents from small data sets,'' \emph{Physica D}, vol.~65, no.~1--2, pp. 117--134, 1993.

\bibitem{hurst1951} H.~E. Hurst, ``Long-term storage capacity of reservoirs,'' \emph{Trans. Am. Soc. Civ. Eng.}, vol.~116, pp. 770--799, 1951.

\bibitem{higuchi1988} M. Higuchi, ``Approach to an irregular time series on the basis of the fractal theory,'' \emph{Physica D}, vol.~31, no.~2, pp. 277--283, 1988.

\bibitem{katz1988} E.~S. Katz, ``Fractal dimension of a time‐varying signal,'' \emph{Fractals}, vol.~1, no.~1, pp. 19--87, 1988.

\bibitem{lempel1976} A. Lempel and J. Ziv, ``On the complexity of finite sequences,'' \emph{IEEE Trans. Inf. Theory}, vol.~22, no.~1, pp. 75--81, 1976.

\bibitem{shannon1948} C.~E. Shannon, ``A mathematical theory of communication,'' \emph{Bell Syst. Tech. J.}, vol.~27, pp. 379--423, 1948.

\bibitem{bandt2002} C. Bandt and B. Pompe, ``Permutation entropy: a natural complexity measure for time series,'' \emph{Phys. Rev. Lett.}, vol.~88, no.~17, Art. 174102, 2002.

\bibitem{richman2000} J.~S. Richman and J.~R. Moorman, ``Physiological time-series analysis using approximate entropy and sample entropy,'' \emph{Am. J. Physiol. Heart Circ. Physiol.}, vol.~278, no.~6, pp. H2039--H2049, 2000.

\bibitem{pincus1991} S.~M. Pincus, ``Approximate entropy as a measure of system complexity,'' \emph{Proc. Natl. Acad. Sci. USA}, vol.~88, no.~6, pp. 2297--2301, 1991.

\bibitem{tsallis1988} C. Tsallis, ``Possible generalization of Boltzmann–Gibbs statistics,'' \emph{J. Stat. Phys.}, vol.~52, no.~1--2, pp. 479--487, 1988.

\bibitem{schiff2009} S.~J. Schiff, T. Sauer, R. Kumar, and S.~L. Weinstein, ``Neuronal spatiotemporal pattern discrimination: The dynamical evolution of seizure and spreading depression,'' \emph{Med. Biol. Eng. Comput.}, vol.~47, no.~5, pp. 419--428, 2009.

\end{thebibliography}
\end{document}